\documentclass[12pt,a4paper]{article}
\usepackage[leqno]{amsmath}
\usepackage{graphicx}
\title{Model of a motion of substance in a channel of a network consisting of two arms}
\author{Nikolay K. Vitanov\footnote{corresponding author, e-mail:vitanov@imbm.bas.bg}, Roumen Borisov}
\date{Institute of Mechanics, Bulgarian Academy of Sciences, Akad. G. Bonchev Str., Bl. 4, 1113 Sofia, Bulgaria}

\begin{document}
\maketitle
\begin{abstract}
We study the problem of the motion of substance in a channel of a network
for the case of channel having two arms. Stationary regime of the flow of 
the substance is considered. Analytical relationships for the distribution 
of the substance in the nodes of the arms of the
channel are obtained. The obtained results are discussed from the point of
view of technological applications of the model (e.g., motion of substances 
such as water in complex technological facilities).
\end{abstract}
\section{Introduction}
In the last decades the researchers realized the importance of  dynamics 
of complex systems and this lead to intensive studies of such systems, especially 
in the area of social dynamics and population dynamics \cite{cs1} - \cite{cs5}. 
In the course of this research the networks have appeared as important part of 
the structure of many complex systems \cite{cs6} - \cite{cs8}. And an important 
part of the processes in a network are the network flows.
Research on network flows  has many  roots and some of them are in the studies on 
transportation  problems \cite{ff} or in the studies on migration flows \cite{mf1} - \cite{v1}. 
At the beginning of the research the problems of interest have been, e.g., minimal cost 
flow problems or possible maximal flows in a network. Today one uses 
the methodology from the theory of network flows \cite{ch1} to solve 
problems connected to ,e.g.,  just in time scheduling,  facility 
layout and location or electronic route guidance in urban traffic 
networks \cite{hani}. 
\par
Below we shall consider a network consisting of nodes and edges - Fig. 1.
Some of the nodes of the network (pictured by circles in Fig.1)
and the corresponding edges (represented by solid lines in Fig. 1)
form a channel. This channel has a single arm  up to the node $n$ where
the channel splits to two arms. We shall consider a flow of substance 
through such a channel. For an example the  channel  can consists of  \
chains of  cells connected by transportation systems that
transport the substance from one cell to the next cell of the channel. 
We can think about  such a situation as a situation reflecting the case 
of logistic channel  consisting of storing facilities and each of these 
facilities supplies a city from a network of cities. Another possible
interpretation of such a channel is motion of some substance (e.g. water)
through cells of complex technological system. In this case the edges
can be water pipes. 
\par
The following processes can be observed 
in a node of the studied channel:  exchange (inflow and outflow) of 
substance with the previous node of the 
channel;  exchange (outflow and inflow) of substance with the next node of the channel;
exchange (inflow and outflow) of substance with the environment;
"leakages":   outflow and inflow of substance to and from the corresponding
node of the network. 
Below we shall consider the stationary regime of the functioning of the
channel (i.e., the regime where the amount of the substance in the cells
doesn't depend explicitly on the time: the quantities of the substance that
enter and leave the channel are the same). Our interest will be focused on
distributions of the substance in the nodes of the studied channel for the case
of this stationary regime.
Studied model has numerous applications, e.g, in migration channels, in 
scientometrics, in logistics, etc. \cite{cs4}, \cite{v1}. 
\begin{figure}[!htb]
\centering
\includegraphics[scale=.5]{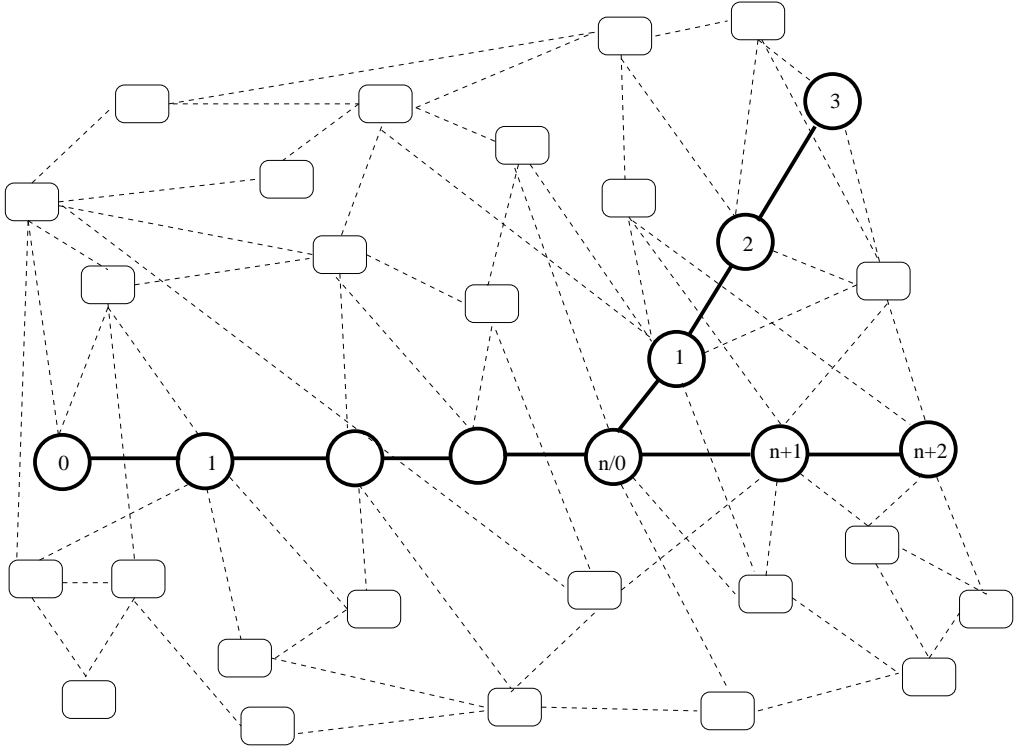}
\caption{Part of a network and the two arms of the studied channel. The nodes 
of the arms are represented by circles and the edges that connect the 
nodes of the arms are represented by solid lines. At a selected node the 
channel splits to two arms. This node is labelled $n/0$. $n$ is the 
number of the splitting node from
the point of view of the first arm of the channel (the horizontal chain
of nodes). $0$ is the number of the node from the point of view of the
second arm of the channel. The  nodes of the network that are not a part of
the studied channel are connected by edges that are represented by dashed 
lines. Note that the entry nodes of the two arms of the channel are labelled
with $0$.}
\end{figure}
\section{Mathematical formulation of the problem}
Specific feature of 
the study below is that the channel splits to two arms at the $n$-th node.
Thus there are two special nodes in our case. The first node of the network
(called also the entry node and labeled by the number $0$) is the only node of 
the network where the substance may enter the channel. 
The second special node of the network is the node where the channel
splits in two arms. We assume that the substance moves only in one direction
along the channel (from nodes labeled by smaller numbers to nodes labeled by
larger numbers). The "leakage" is also only in the direction from the channel to
the network (and not in the opposite direction). 
\par
Let us now consider the situation where the single arm of the channel contains 
$n+1$-nodes (the nodes labeled by the numbers from $0$ to $n$) and then the 
channel splits to two arms. From the point of view of Fig. 1 this is the part
of the channel that is shown in the left-hand side of the picture before the
node where the channel splits to two arms.  This part of the channel
consists of a chain of nodes of a network. The nodes are
connected by edges and each node is connected only to the two neighboring 
nodes of the channel exclusive for the first  node of the channel that is 
connected only to the neighboring node. We study a model of the 
motion of substance through such a
channel which is an extension of the model discussed in  \cite{sg1} and
\cite{v1}. We consider each node as a cell (box), 
i.e.,  we consider an  array of infinite  number of cells indexed in 
succession by non-negative integers.  
We assume that an amount $x$ of some substance  is
distributed among the cells and this substance can move from one cell to another cell. Let $x_i$ be the amount of the substance in the $i$-th cell. Then
\begin{equation}\label{warigx1}
x = \sum \limits_{i=0}^\infty x_ i
\end{equation}
The fractions $y_i = x_i/x$ can be considered as probability values of
distribution of a discrete random variable $\zeta$
\begin{equation}\label{warigx2}
y_i = p(\zeta = i), \ i=0,1, \dots
\end{equation}
The content $x_i$ of any cell may change as consequence of 
the following 3 processes:
\begin{enumerate}
\item Some amount $s$ of the substance $x$  enters the system of
cells from the external environment through the $0$-th cell;
\item Rate $f_i$ from $x_i$ is transferred from the $i$-th
cell into the $i+1$-th cell;
\item Rate $g_i$ from  $x_i$  leaks out the $i$-th cell into the
external environment.
\end{enumerate}
We assume that the process of the motion of the substance is continuous in the
time. Then the process can be modeled mathematically by the system of ordinary
differential equations:
\begin{eqnarray} \label{warigx4}
\frac{dx_0}{dt} &=& s-f_0-g_0; \nonumber \\
\frac{dx_i}{dt} &=& f_{i-1} -f_i - g_i, \ i=1,2,\dots.
\end{eqnarray}
\par
There are  two regimes of functioning of the channel: stationary regime 
and non-stationary regime. What we shall discuss below is the stationary regime
of functioning of the channel.
In the stationary regime of the functioning of the channel 
$\frac{dx_i}{dt}=0$, $i=0,1,\dots$. Let us mark the quantities for the stationary case with 
$^*$. Then from Eqs.(\ref{warigx4}) one obtains 
\begin{equation}\label{st1}
f_0^*=s^*-g_0^*; \ \  f_i^*=f_{i-1}^*-g_i.
\end{equation}  
This result can be written also as
\begin{equation}\label{st2}
f_i^* = s^*- \sum \limits_{j=0}^i g_j^*
\end{equation}
Hence for the stationary case the situation in the channel is determined by
the quantities $s^*$ and $g_j^*$, $j=0,1,\dots$. 
In this paper we shall assume the following forms of the amount 
of the moving substances  in 
Eqs.(\ref{warigx4}) ($\alpha, \beta, \gamma_i, \sigma$ are constants)
\begin{eqnarray}\label{warigx5}
s &=& \sigma_0 x_0  \ \ \sigma_0 > 0  \nonumber \\
f_i &=& (\alpha_i + \beta_i i) x_i; \ \ \ \alpha_i >0, \ \beta_i \ge 0  \nonumber \\
g_i &=& \gamma^*_i x_i; \ \ \ \gamma^*_i \ge 0 \to \textrm{non-uniform leakage 
in the nodes}
\end{eqnarray}
$\gamma^*_i$ is a quantity specific for the present study. 
$\gamma^*_i = \gamma_i + \delta_i $ describes the situation with the
leakages in the nodes of the channel. We shall assume that $\delta_i=0$
for all $i$ except for $i=n$. This means that in the $n$-th node (where the
second arm of the channel splits from the first arm of the channel) in addition
to the usual leakage $\gamma_i$ there is additional leakage of substance 
given by the term $\delta_n x_n$ and this additional leakage supplies the
substance that then begins its motion along the second arm of the channel. 
\par
On the basis of all above the model system of differential equations for 
this arm of the channel becomes
\begin{eqnarray} \label{eq1}
\frac{dx_{0}}{dt}&=&\sigma_0 x_0-\alpha_0 x_0-\gamma^*_0 x_0  \nonumber \\
\frac{dx_{i}}{dt}&=&[\alpha_{i-1}+(i-1)\beta_{i-1}]x_{i-1}-(\alpha_{i}+ i \beta_{i}+\gamma^*_{i})x_{i};\ \ i=1,2,\dots 
\end{eqnarray}
\par
We shall consider the stationary regime of functioning of the channel. Then
$dx_0/dt=0$ and
from the first of the Eqs.(\ref{eq1}) it follows that
$\sigma_0 = \alpha_0 + \gamma_0$. This means that $x_0$
(the amount of the substance in the $0$-th cell of the channel) is free
parameter. In principle the solution of Eqs.(\ref{eq1}), $i=1,2,\dots$ is
\begin{equation}\label{eq2}
x_i = x_i^* + \sum \limits_{j=0}^i b_{ij} \exp[-(\alpha_j + j \beta_j + \gamma^*_j)t]
\end{equation}
where $x_i^*$ is the stationary part of the solution. For $x_i^*$ one obtains
the relationship (just set dx/dt = 0 in the second of Eqs.(\ref{eq1}))
\begin{equation}\label{eq3}
x_i^* = \frac{\alpha_{i-1} + (i-1) \beta_{i-1}}{\alpha_i + i \beta_i + \gamma^*_i} x_{i-1}^*, \ i=1,2,\dots
\end{equation}
The corresponding relationships for the coefficients $b_{ij}$ are
($i=1,\dots$):
\begin{equation}\label{eq4}
b_{ij} = \frac{\alpha_{i-1} + (i-1) \beta_{i-1}}{(\alpha_i - \alpha_j) + (i \beta_i -
j \beta_j) + (\gamma^*_i - \gamma^*_j)} b_{i-1,j},
\ j=0,1,\dots,i-1
\end{equation}
From Eq.(\ref{eq3}) one obtains
\begin{equation}\label{eq5}
x_i^* = \frac{\prod \limits_{j=0}^{i-1}[\alpha_{i-j-1}+(i-j-1)\beta_{i-j-1}]}{
\prod \limits_{j=0}^{i-1} \alpha_{i-j} + (i-j) \beta_{i-j} + \gamma^*_{i-j}} x_0^*
\end{equation}
The form of the corresponding stationary distribution $y_i^* = x_i^*/x^*$ 
(where $x^*$ is the amount of the substance in all of the cells of the arm of the channel) is
\begin{equation}\label{eq6}
y_i^* = \frac{\prod \limits_{j=0}^{i-1}[\alpha_{i-j-1}+(i-j-1)\beta_{i-j-1}]}{
\prod \limits_{j=0}^{i-1} [\alpha_{i-j} + (i-j) \beta_{i-j} + \gamma^*_{i-j}]} y_0^*
\end{equation}
To the best of our knowledge the distribution presented by
Eq.(\ref{eq6}) was not discussed up to now outside our research group. 
Let us show that this
distribution contains as particular cases several famous distributions, e.g.,
Waring distribution, Zipf distribution, and Yule-Simon distribution. In order to
do this we consider the particular case when $\beta_i \ne 0$ and write $x_i$
from Eq.(\ref{eq5}) as follows
\begin{equation}\label{eq7}
x_i^* = \frac{\prod \limits_{j=0}^{i-1} \tilde{b}_{i-j} [k_{i-j-1} + (i-j-1)]}{\prod \limits_{j=0}^{i-1} [k_{i-j} + a_{i-j} + (i-j)]} x_0^*
\end{equation}
where $k_i = \alpha_i/\beta_i$; $a_i = \gamma^*_i/\beta_i$; 
$\tilde{b}_i = \beta_{i-1}/\beta_i$.
The form of the corresponding stationary distribution $y_i^* = x_i^*/x^*$ is
\begin{equation}\label{eq8}
y_i^* = \frac{\prod \limits_{j=0}^{i-1}\tilde{ b}_{i-j} [k_{i-j-1} + (i-j-1)]}{\prod \limits_{j=0}^{i-1} [k_{i-j} + a_{i-j} + (i-j)]} y_0^*
\end{equation}
Let us now consider the particular case where $\alpha_i = \alpha$ and 
$\beta_i = \beta$ for $i=0,1,2,\dots$. Then from Eqs.(\ref{eq7}) and (\ref{eq8}) one obtains
\begin{equation}\label{eq9}
x_i^* = \frac{[k+(i-1)]!}{(k-1)! \prod \limits_{j=1}^i (k+j+a_j)} x_0^*
\end{equation}
where $k = \alpha/\beta$ and $a_j=\gamma^*_j/\beta$.
The form of the corresponding stationary distribution $y_i^* = x_i^*/x^*$ is
\begin{equation}\label{eq10}
y_i^* = \frac{[k+(i-1)]!}{(k-1)! \prod \limits_{j=1}^i (k+j+a_j)} y_0^*
\end{equation}
Let us consider the particular case where $a_0 = \dots = a_N$. In this case the
distribution from Eq.(\ref{eq10}) is reduced to the distribution:
\begin{eqnarray}\label{eq11}
P(\zeta = i) &=& P(\zeta=0) \frac{(k-1)^{[i]}}{(a+k)^{[i]}}; \ \ k^{[i]} = \frac{(k+i)!}{k!}; \ i=1, 2, \dots 
\end{eqnarray}
$P(\zeta=0)=y_0^* = x_0^*/x^*$ is the percentage of substance that is located in
the first cell of the channel. Let this percentage be 
\begin{equation}\label{eq12}
y_0^* = \frac{a}{a+k}
\end{equation}
The case described by Eq.(\ref{eq12}) corresponds to the situation where the
amount of substance in the first cell is proportional of the amount of substance
in the entire channel. In this
case Eq.(\ref{eq10}) is reduced to:
\begin{eqnarray}\label{eq13}
P(\zeta = i) &=& \frac{a}{a+k} \frac{(k-1)^{[i]}}{(a+k)^{[i]}}; \ \ k^{[i]} = \frac{(k+i)!}{k!}; \ i=1, 2, \dots 
\end{eqnarray}
The distribution (\ref{eq13}) is exactly the Waring distribution (probability distribution of non-negative 
integers named after Edward Waring - the 6th Lucasian professor of Mathematics 
in Cambridge from the 18th century)  \cite{varyu1} - \cite{varyu3}.
The mean $\mu$ (the expected value)  and the variance $V$ of the Waring distribution are
\begin{equation}\label{eq16}
\mu = \frac{k}{a -1} \ \textrm{if} \ a >1; \ \
V = \frac{k a  (k + a -1)}{(a-1)^2(a - 2)} \
\textrm{if} \ a >2
\end{equation}
$\rho$ is called the tail parameter as it controls the tail of the Waring
distribution. Waring distribution contains various distributions as particular cases. Let $i \to \infty$ Then the Waring distribution is reduced to  the frequency form of the Zipf distribution \cite{chen}
\begin{equation}\label{eq18}
P(\zeta=i) \approx \frac{1}{i^{(1+ a)}}.
\end{equation}
If $k \to 0$ the Waring distribution is reduced to the Yule-Simon distribution \cite{simon} 
\begin{equation}\label{eq19}
P(\zeta = i ) = a B(a+1,i)
\end{equation}
where $B$ is the beta-function. 
\par
Let us now consider the stationary regime of functioning of the second arm
of the channel. Here we shall denote as $0$-th node the node where the
second arm of the channel splits from the first arm. 
We assume that an amount $z$ of the substance  becomes distributed among the
 cells of the second arm of the channel and this substance can move from one cell to another cell. Let $z_i$ be the amount of the substance in the $i$-th cell. Then
\begin{equation}\label{warigx1x}
z = \sum \limits_{i=0}^\infty z_i
\end{equation}
The fractions $y_i = z_i/z$ can be considered as probability values of
distribution of a discrete random variable $\zeta$
\begin{equation}\label{warigx2x}
y_i = p(\zeta = i), \ i=0,1, \dots
\end{equation}
The content $z_i$ of any cell may change due to the same three processes
that govern the motion of the substance in the first arm of the channel
The process of the motion of the substance is continuous in the
time. Then the process can be modeled mathematically by the system of ordinary
differential equations:
\begin{eqnarray} \label{warigx4x}
\frac{dz_0}{dt} &=& \hat{s}-\hat{f}_0-\hat{g}_0; \nonumber \\
\frac{dz_i}{dt} &=& \hat{f}_{i-1} - \hat{f}_i - \hat{g}_i, \ i=1,2,\dots.
\end{eqnarray}
The  relationships for the quantities of the above equations are 
($\hat{\alpha}, \hat{\beta}, \hat{\gamma_i}$ are constants)
\begin{eqnarray}\label{warigx5x}
\hat{s} &=& \delta_n x_n;   \nonumber \\
\hat{f}_i &=& (\hat{\alpha}_i + \hat{\beta}_i i) x_i; \ \ \ \hat{\alpha}_i >0, \ \hat{\beta}_i \ge 0  \nonumber \\
\hat{g}_i &=& \hat{\gamma}_i x_i; \ \ \ \hat{\gamma}_i \ge 0 
\end{eqnarray}
Thus the system of equations for the motion of the substance in
this arm of the channel is
\begin{eqnarray} \label{eq20}
\frac{dz_{0}}{dt}&=&\delta_n x_n-\hat{\alpha}_0 z_0- \hat{\gamma}_0 z_0  \nonumber \\
\frac{dz_{i}}{dt}&=&[\hat{\alpha}_{i-1}+(i-1)\hat{\beta}_{i-1}]z_{i-1}-
(\hat{\alpha}_{i}+ i \hat{\beta}_{i}+\hat{\gamma}_{i})z_{i};\ \ i=1,2,\dots 
\end{eqnarray}
In this article we shall discuss the situation in which the stationary
state is established in the entire channel (in the two arm of the channel).
In this case $x_n \to x_n^*$; $\frac{dz_{0}}{dt} \to 0$ and $\frac{dz_{i}}{dt} \to 0$. Then
\begin{eqnarray}\label{eq21}
z_0^* = \frac{\delta_n x_n^*}{\hat{\alpha}_0 + \hat{\gamma_0}}; \ \
z_i^* = \frac{\hat{\alpha}_{i-1} + (i-1) \hat{\beta}_{i-1}}{\hat{\alpha}_i + i \hat{\beta}_i + \hat{\gamma}_i} z_{i-1}^*, \ i=1,2,\dots
\end{eqnarray} 
From the second of Eqs.(\ref{eq21}) one obtains
\begin{equation}\label{eq22}
z_i^* = \frac{\prod \limits_{j=0}^{i-1}[\hat{\alpha}_{i-j-1}+(i-j-1)\hat{\beta}_{i-j-1}]}{
\prod \limits_{j=0}^{i-1} \hat{\alpha}_{i-j} + (i-j) \hat{\beta}_{i-j} + \hat{\gamma}_{i-j}} z_0^*
\end{equation}
The form of the corresponding stationary distribution $y_i^* = z_i^*/z^*$ 
(where $z^*$ is the amount of the substance in all of the cells of second  
arm of the channel) is
\begin{eqnarray}\label{eq23}
y_0^* = 
\frac{1}{1+\sum \limits_{i=1}^\infty \frac{\prod \limits_{j=0}^{i-1}
		[\hat{\alpha}_{i-j-1}+(i-j-1)\hat{\beta}_{i-j-1}]}{
		\prod \limits_{j=0}^{i-1} \hat{\alpha}_{i-j} + (i-j) \hat{\beta}_{i-j} + \hat{\gamma}_{i-j}}}; \ \
y_i^* = 
\frac{
\frac{\prod \limits_{j=0}^{i-1}[\hat{\alpha}_{i-j-1}+(i-j-1)\hat{\beta}_{i-j-1}]}{
\prod \limits_{j=0}^{i-1} \hat{\alpha}_{i-j} + (i-j) \hat{\beta}_{i-j} + 
\hat{\gamma}_{i-j}}}{1+\sum \limits_{i=1}^\infty \frac{\prod \limits_{j=0}^{i-1}
[\hat{\alpha}_{i-j-1}+(i-j-1)\hat{\beta}_{i-j-1}]}{
\prod \limits_{j=0}^{i-1} \hat{\alpha}_{i-j} + (i-j) \hat{\beta}_{i-j} + \hat{\gamma}_{i-j}}}, \nonumber \\
 i=1,2,\dots
\end{eqnarray}
\par
On the basis of the analogy between Eqs. (\ref{eq6}) and (\ref{eq22}) one can easily
see that the Waring distribution is a particular case also for the distribution given by
Eq.(\ref{eq23}) that describes the distribution of the substance in the second arm of the
channel. One has just to repeat the calculations starting from Eq.(\ref{eq7}) and finishing at
Eq.(\ref{eq19}).
\section{Concluding remarks}
In this article we  obtain analytical relationships for the distribution of the
substance in the two arm of a channel of a network for the case of the stationary 
regime of the functioning of the channel. On the basis of these relationships 
we can make numerous conclusions. Let us discuss just one of these conclusions:
\emph{the presence of the second arm of the channel changes the distribution of
the substance in the first arm of the channel}. In order to discuss this let us
denote as $y_i^{*(1)}$ the distribution of the substance in the cells of the
first arm of the channel for the case of lack of second arm of the channel.
Let $y_i^{*(2)}$ be the distribution of the substance in the cells of the
first arm of the channel for the case of presence of second arm of the channel.
From the theory in the previous section one easily obtains the relationship
\begin{eqnarray}\label{eq31}
\frac{y_i^{*(1)}}{y_i^{*(2)}} = 
\prod \limits _{j=0}^{i-1} 
\frac{\alpha_{i-j} + (i-j) \beta_{i-j} + \gamma^*_{i-j}}{\alpha_{i-j} + (i-j) \beta_{i-j} + \gamma_{i-j}} = \prod \limits _{j=0}^{i-1}
\left[ 1+ \frac{\delta_{i-j}}{\alpha_{i-j} + (i-j) \beta_{i-j} + \gamma_{i-j}}\right] \nonumber \\
\end{eqnarray}
When $i < n$ then $\delta_i=0$ and there is no difference between the
distribution of the substances in the channel with single arm and in the channel with two arms. The difference arises at the splitting cell (the $n$-th
cell from the point of view of the numbering of the first arm of the channel).
As it can be easily calculated for $i \ge n$ Eq.(\ref{eq31}) reduces to
\begin{eqnarray}\label{eq32}
\frac{y_i^{*(2)}}{y_i^{*(1)}} = \frac{1}{ 
1+ \frac{\delta_{n}}{\alpha_{n} + n \beta_{n} + \gamma_{n}}}, \ \ i \ge n
\end{eqnarray}
Eq.(\ref{eq32}) shows clearly that splitting of the channel affects the tail
of the distribution of the substance in the first arm of the channel.
If the second arm of the channel don't exists then $\delta_n=0$ and the
distribution of the substance is a long-tail  distribution
that contains the long-tail  Waring distribution as a particular
case. The "leakage" of the substance to second arm of the channel may reduce
much the tail of the distribution of substance in the first arm of the channel. This reduction can even be kink-wise at the $n$-th node for large value of $\delta_n$.  

\end{document}